\documentclass[12pt]{article}
\usepackage{graphicx}

\def\pbnr{}
\def\speaker{Jonna Koponen}
\def\onbehalfof{}
\def\title{Lattice results for $D/D_s$ leptonic and semileptonic decays}
\def\affiliation{SUPA, School of Physics and Astronomy,\\
University of Glasgow, Glasgow, G12 8QQ, UK}
\def\support{}

\newcommand\pubnumber{\pbnr}
\newcommand\pubdate{\today}
\def\Title#1{\begin{center} {\Large #1 } \end{center}}
\def\Author#1{\begin{center}{ \sc #1} \end{center}}

\newcommand{\OnBehalf}[1]{\sbox0{#1}\ifdim\wd0=0pt
        {}
	\else
	{\\on behalf of #1}
	\fi}
\newcommand{\SupportedBy}[1]{\sbox0{#1}\ifdim\wd0=0pt
        {}
	\else
	{\footnote{#1}}
	\fi}
\def\Address#1{\begin{center}{ \it #1} \end{center}}

\newcommand\pubblock{\includegraphics[width=5cm]{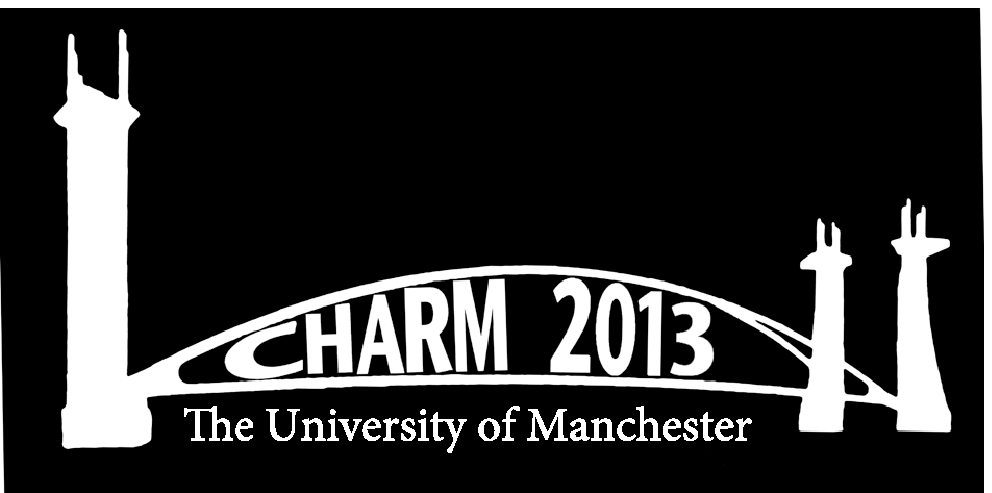}\hfill{\begin{tabular}{l} \pubnumber\\
         \pubdate  \end{tabular}}}
\newenvironment{Abstract}{\begin{quotation}  }{\end{quotation}}
\newenvironment{Presented}{\begin{quotation} \begin{center} 
             PRESENTED AT\end{center}\bigskip 
      \begin{center}\begin{large}}{\end{large}\end{center} \end{quotation}}
\def\Acknowledgements{\bigskip  \bigskip \begin{center} \begin{large}
             \bf ACKNOWLEDGEMENTS \end{large}\end{center}}
\def\venue{The 6$^{th}$ International Workshop on Charm Physics\\
(CHARM 2013)\\
Manchester, UK,  31 August -- 4 September, 2013}

\def\beq{\begin{equation}}
\def\eeq#1{\label{#1}\end{equation}}
\def\eeqn{\end{equation}}

\def\beqa{\begin{eqnarray}}
\def\eeqa#1{\label{#1}\end{eqnarray}}
\def\eeqan{\end{eqnarray}}

\let\bar=\overbar

\def\Dslash{\not{\hbox{\kern-4pt $D$}}}
\def\dslash{\not{\hbox{\kern-2pt $\del$}}}

\def\msb{{\bar{\ssstyle M \kern -1pt S}}}

\begin{document}
\begin{titlepage}
\pubblock

\vfill
\Title{\title}
\vfill
\Author{\speaker\SupportedBy{\support}\OnBehalf{\onbehalfof}}
\Address{\affiliation}
\vfill
\begin{Abstract}
This review article summarizes recent lattice QCD results for $D$
and $D_s$ meson leptonic and semileptonic decays. Knowing the meson
decay constants and semileptonic form factors from theory, one can
extract CKM elements $V_{cd}$ and $V_{cs}$ from experimental results.
At present, the most accurate results for decay constants are from the
Fermilab Lattice and MILC Collaborations \cite{FNAL_MILC13}:
$f_D = 212.5 \pm 0.5_{\mathrm{stat}} {}^{+0.6}_{-1.5} |_{\mathrm{syst}}$~MeV and
$f_{D_s} = 248.9 \pm 0.2_{\mathrm{stat}} {}^{+0.5}_{-1.6}|_{\mathrm{syst}}$~MeV,
giving
$V_{cd} = 0.2184 \pm 0.009_{\mathrm{expt}} {}^{+0.0008}_{-0.0016} |_{\mathrm{lattice}}$
and $V_{cs} = 1.017 \pm 0.02_{\mathrm{expt}} {}^{+0.002}_{-0.007} |_{\mathrm{lattice}}$.
The shapes of the semileptonic form factors from lattice QCD agree very
well with experiment, and the accuracy is currently at the $2-5$\%
level for $D \to \pi\ell\nu$ and $1-2$\% for $D \to K\ell\nu$.
Extracting the CKM elements from the semileptonic decays yields
$V_{cd} = 0.225(6)_{\mathrm{expt}}(10)_{\mathrm{lattice}}$
(HPQCD Collaboration, from~\cite{HPQCD11}) and
$V_{cs} = 0.963(5)_{\mathrm{expt}}(14)_{\mathrm{lattice}}$
(HPQCD Collaboration, from~\cite{HPQCD13}).
These lattice calculations also revealed that the semileptonic form
factors are insensitive to whether the spectator quark is a light or
strange quark.
\end{Abstract}
\vfill
\begin{Presented}
\venue
\end{Presented}
\vfill
\end{titlepage}
\def\thefootnote{\fnsymbol{footnote}}
\setcounter{footnote}{0}
%

\section{Motivation}

$D$ and $D_{s}$ meson decays are a very interesting research area at the moment.
On the experimental side, BES III and Belle have presented preliminary results
from their recent runs; on the theory side, lattice QCD is able to provide
non-perturbative, precise calculations of meson decay constants and semileptonic
form factors from first principles. Combining the experimental and theoretical
results allows us to determine elements $|V_{cs}|$ and $|V_{cd}|$ of the quark mixing
matrix (the CKM matrix). Several processes can be used to extract the same CKM
matrix element, which allows for cross checks and consistency tests of the Standard
Model and constraints/test for new physics. Similar methods can be  used to study
$B$ and $B_s$ meson decays, so charm decays are an excellent test environment for these
lattice QCD tools.

The aim of this review is to summarize recent lattice QCD results for the leptonic
and semileptonic decays, i.e. decay constants $f_D$ and $f_{D_s}$, and form factors
for $D \to K\ell\nu$ and $D \to \pi\ell\nu$. The article is divided into three parts:
a general introduction, lattice results and CKM elements $V_{cd}$ and $V_{cs}$.

\section{Introduction}

\subsection{Leptonic and semileptonic decays}

In a leptonic decay a meson (here $D$ or $D_s$) decays to a lepton and its neutrino via a 
virtual $W$ boson. The decay rate is given by
\begin{equation}
\Gamma^{D_s \to \ell\nu} = \frac{G_F^2}{8\pi}m_\ell^2M_{D_s}
\Bigg(1-\frac{m_\ell^2}{M_{D_s}^2}\Bigg)^2f^2_{D_s}|V_{cs}|^2.
\end{equation}
Hence $f^2_{D_s}|V_{cs}|^2$ can be cleanly extracted from experiment. The decay constant
$f_{D_s}$ (or $f_D$ for a $D$ meson decay) is a property of the hadron
and can thus be calculated in lattice QCD.

On the other hand, consider a semileptonic decay where a $D$ meson decays to a $K$
meson (or a pion), a lepton and its neutrino via a virtual $W$ boson. If both the
initial and final state mesons are pseudoscalars, the partial decay rate can be
written as
\begin{equation}
\frac{\mathrm{d}\Gamma^{D \to K}}{\mathrm{d}q^2}=\frac{G_F^2p^3}{24\pi^3}
|V_{cs}|^2|f^{D \to K}_+(q^2)|^2.
\end{equation}
Here $p=|\vec{p}|$ is the momentum of the $K$ meson in the rest frame of the $D$,
and $q^2$ is the four-momentum transfer between the two mesons,
\begin{equation}
q^2=(M_{D}-E_K)^2-p^2.
\end{equation}
Again, experiment can tell us $|V_{cs}|^2|f^{D \to K}_+(q^2)|^2$
as a function of $q^2$. The form factor $f_+$ is a QCD quantity, and again
this can be determined in a lattice QCD calculation. Note that the same CKM element
appears in both cases.

\subsection{Lattice QCD}

At the moment lattice QCD is the only known method that can provide a precise,
non-perturbative theoretical determination of form factors and decay
constants. In a lattice calculation space-time is discretized to make a 4
dimensional box with lattice spacing $a$ to allow numerical
integration of the QCD path integral.
There are many details involved in such a calculation, but a rough
sketch of one would be:
\begin{enumerate}
\item 
Generate sets of gluon fields for Monte Carlo integration of the path integral
(including effects of sea quarks).
\item 
Calculate averaged ``hadron correlators'' from valence quark
propagators calculated on these gluon fields.
\item
Fit the correlators as a function of time to obtain masses and simple matrix
elements.
\item
Determine lattice spacing $a$ and fix quark masses using experimental information
(often meson masses) to get results in physical units.
\item 
Extrapolate to $a=0$ and the physical $u/d$ quark mass for real world.
\end{enumerate}
Lattice calculations have had to work with heavier than physical $u/d$ masses
because of numerical cost. Lattices with physical $m_{u,d}$ are now being generated
and the extrapolation to physical light quark masses is becoming just a small
correction.

\section{Lattice results}

\subsection{Leptonic decays, decay constants}
\label{decayconst}

\begin{figure}
\centering
\includegraphics[width=0.99\textwidth]{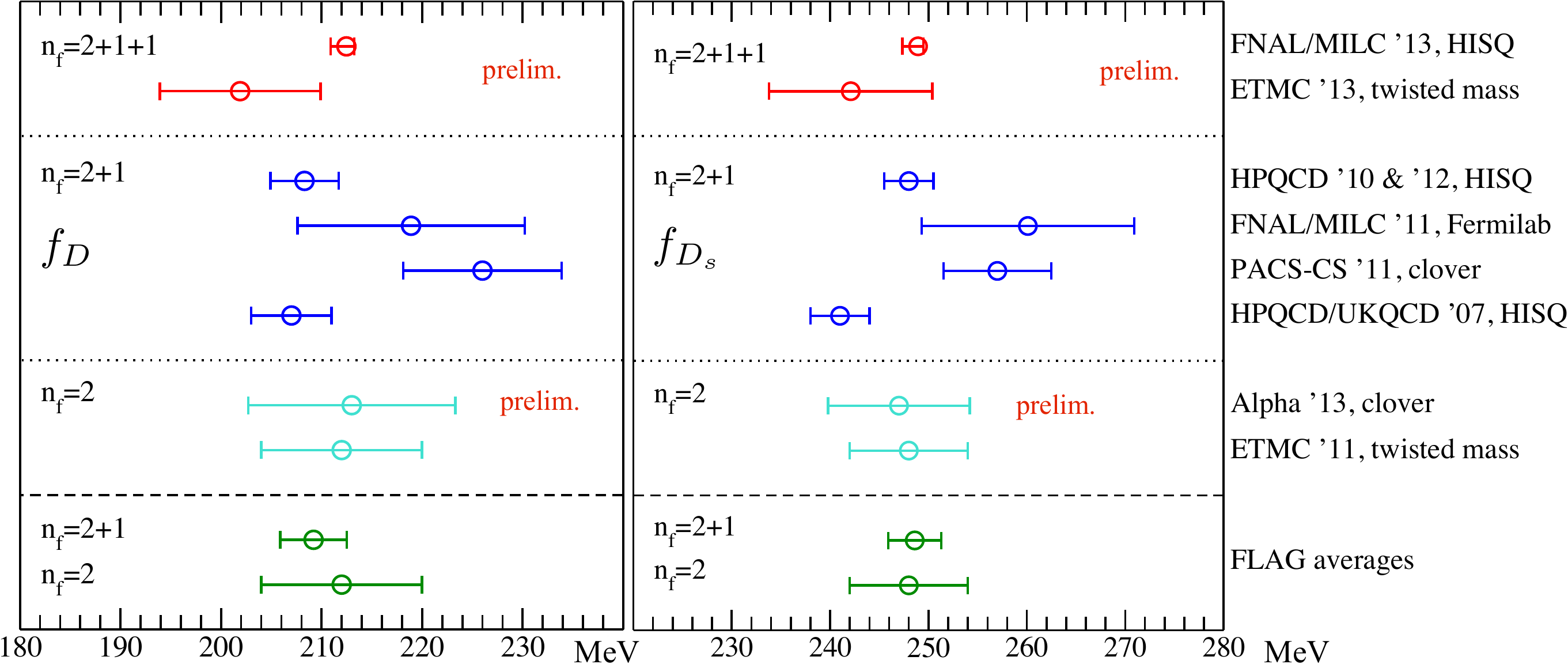}
\caption{Decay constants: $f_D$ on the left, $f_{D_s}$ on the right.
The best results at the moment, i.e. results with
smallest errors and most modern lattice configurations ($n_f=2+1+1$,
physical pion mass), are from
Fermilab Lattice and MILC Collaborations (FNAL/MILC '13):
$f_D = 212.5 \pm 0.5_{\mathrm{stat}} {}^{+0.6}_{-1.5} |_{\mathrm{syst}}$~MeV and
$f_{D_s} = 248.9 \pm 0.2_{\mathrm{stat}} {}^{+0.5}_{-1.6} |_{\mathrm{syst}}$~MeV.}
\label{fig:fDfDs}
\end{figure}

\begin{figure}
\centering
\includegraphics[width=0.7\textwidth]{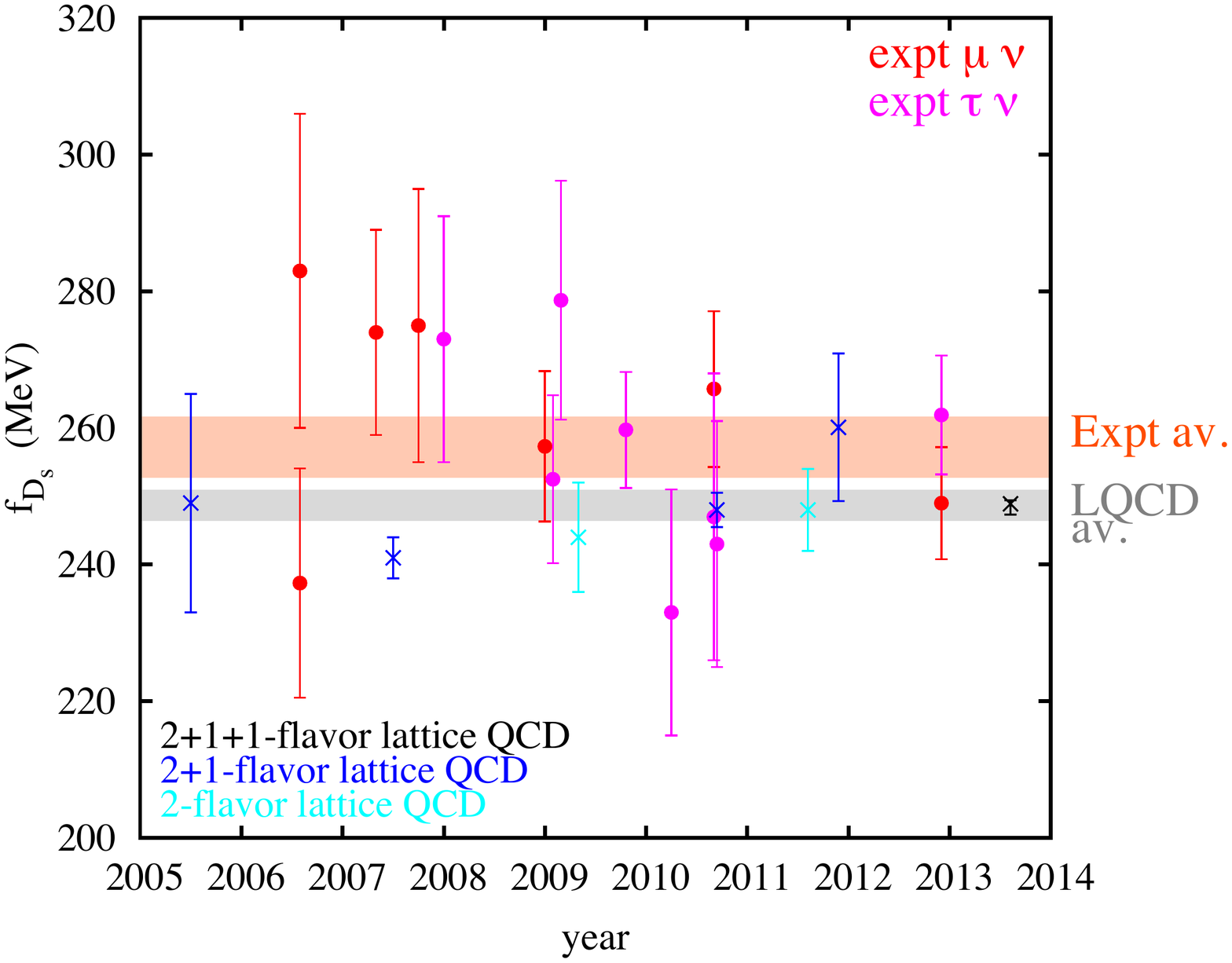}
\caption{The history of $f_{D_s}$. The experimental values have been obtained using
the unitarity value for $|V_{cs}|$ from the PDG (i.e. unitarity of the CKM matrix is
assumed). The darker red data points are for the decay channel $D_s \to \mu\nu$ and the lighter
red for $D_s \to \tau\nu$. Light blue, blue and black crosses are values from $2$
flavor, $2+1$ flavor and $2+1+1$ flavor lattice QCD,
respectively. This figure is an update of Fig. 18 in~\cite{HPQCD10}.}
\label{fig:fDshistory}
\end{figure}

The current status of calculations of the $D$ and $D_s$ meson decay constants
is shown in Fig.~\ref{fig:fDfDs}, tagged by the names of the
lattice groups. The results are from~\cite{ETMC11,HPQCD_UKQCD07,HPQCD10,PACS-CS11,
FNAL_MILC11,HPQCD12,ETMC13,Alpha13,FNAL_MILC13}. Note that some of the results
are still preliminary. $n_f$ denotes the number of flavors used in the calculation:
$n_f=2$ is two light quarks in the sea ($u$ and $d$ quarks that both have the same
mass), $n_f=2+1$ is two light plus strange quarks and $n_f=2+1+1$ has in addition
charm quarks in the sea. The tags ``HISQ'', ``twisted mass'', ``Fermilab'' and
``clover'' denote different discretizations of the Dirac equation for quarks. These different
discretizations should all agree in the continuum limit, and as can be seen in the
figures the agreement is good. For completeness, averages from Flavor
Lattice Averaging Group (FLAG)~\cite{FLAG} are also shown.

It is also interesting to look at the history of the $D_s$ meson decay constant
and see how the value has evolved over the years. This is shown in
Fig.~\ref{fig:fDshistory}. A few years ago there was disagreement between the values
from experiment and lattice, but that has now mostly gone away leaving a tension
of $2\sigma$. The very precise (1\%) value from lattice QCD has
been confirmed by two separate groups and looks solid.

\subsection{Semileptonic decays, form factors}

\begin{figure}
\centering
\includegraphics[width=0.99\textwidth]{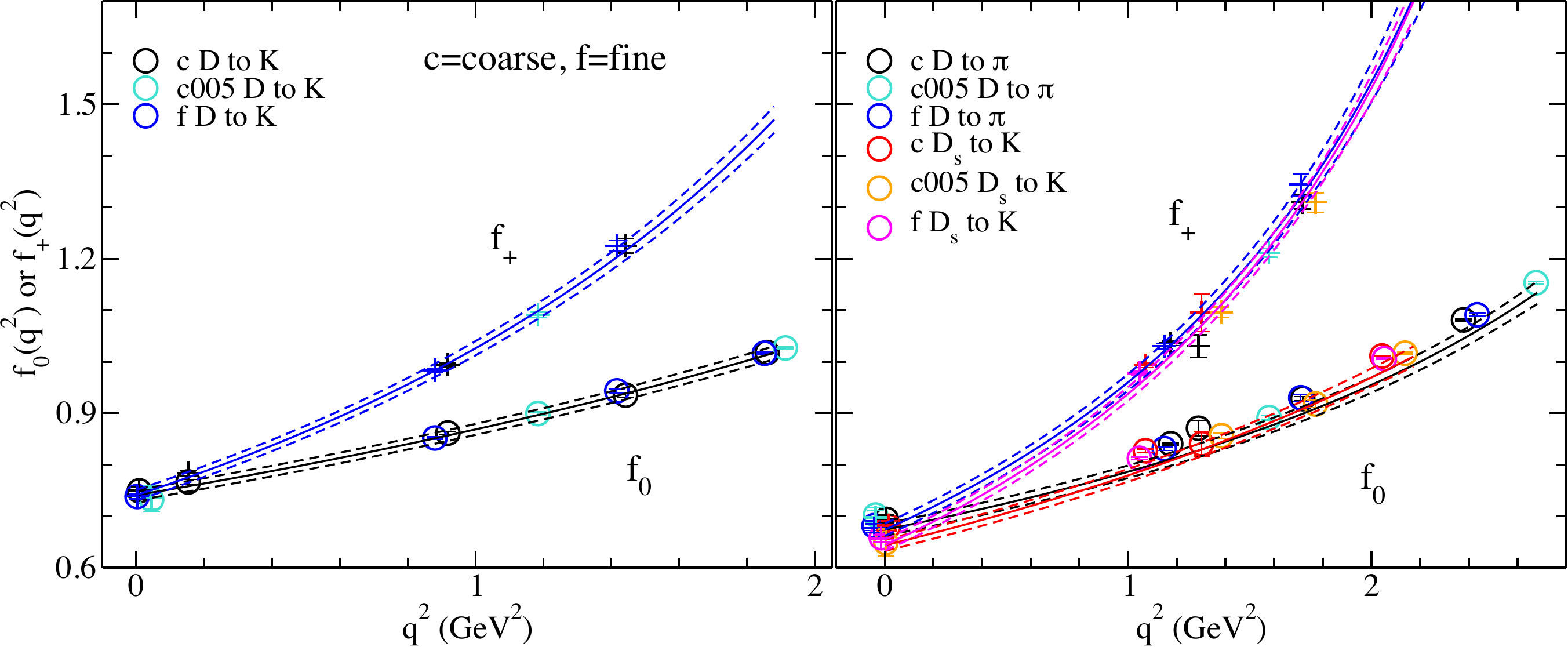}
\caption{On the left: Scalar and vector form factors of $D \to K$
  semileptonic decay~\cite{HPQCD13}. 
On the right: Form factors of $D \to \pi$ and $D_s \to K$ semileptonic decays.
Note that the shape of the form factors is insensitive to the mass
of the spectator quark.}
\label{fig:HPQCDFF}
\end{figure}

Let us turn to $D$ and $D_s$ meson semileptonic decays and their form
factors. In fact there are two form factors, a scalar form factor $f_0$ and 
a vector form factor $f_+$, associated with a pseudoscalar to
pseudoscalar semileptonic decay. In experiment the scalar form factor 
is suppressed by the lepton mass and thus not accessible. However,
on the lattice it is quite straightforward to consider two currents, a 
scalar and a vector current, and calculate both form factors $f_0$ and 
$f_+$. There is also a useful kinematic constraint that $f_+(0)=f_0(0)$.

Here we will only consider lattice results for decays $D \to K\ell\nu$ 
and  $D\to \pi\ell\nu$. Several groups have calculated the shape of 
the $D\to K$ form factors -- see
Refs.~\cite{HPQCD13,FNAL_MILC_DK,Sanfilippo}. Fig.~\ref{fig:HPQCDFF}
shows results by HPQCD from different lattice spacings [coarse ($a=0.12$~fm)
and fine ($a=0.09$~fm) lattice] and extrapolation to continuum and physical
light quark mass (more details of the extrapolation are in Section~\ref{zexpansion}).

The study by HPQCD~\cite{HPQCD13} revealed that the semileptonic decay
form factors are insensitive to spectator quark mass. This is
shown very clearly in Fig.~\ref{fig:HPQCDFF}: the form factors for
$D\to \pi\ell\nu$ and $D_s\to K\ell\nu$ are the same within few
percent, and even within 2\% for most of the $q^2$ range. These decays
are both $c$ to $d$ decays, and the difference is the spectator quark:
a light quark in the $D\to \pi$ case, and strange in $D_s \to K$. This
has been shown to hold for $B \to D\ell\nu$ and $B_s \to D_s\ell\nu$ 
as well~\cite{Bsemilept}.
The same lattice methods can be used to study decays
that involve vector mesons, like weak decay $D_s \to \phi\ell\nu$ or
charmonium radiative decay $J/\psi \to \eta_c\gamma$ -- see for
example~\cite{DsphiJPsigamma}.

\subsection{The z-expansion, continuum and chiral extrapolation}
\label{zexpansion}

\begin{figure}
\centering
\includegraphics[width=0.37\textwidth]{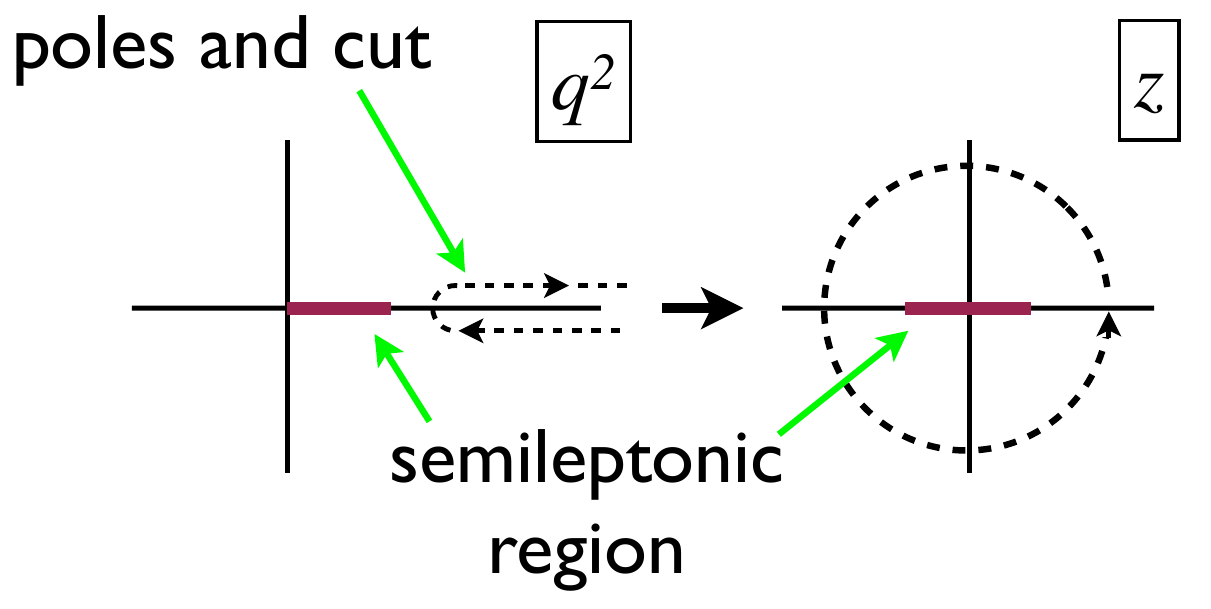}
\caption{Converting the semileptonic region in $q^2$-space to $z$-space.}
\label{fig:q2toz}
\end{figure}

\begin{figure}
\centering
\includegraphics[width=0.65\textwidth]{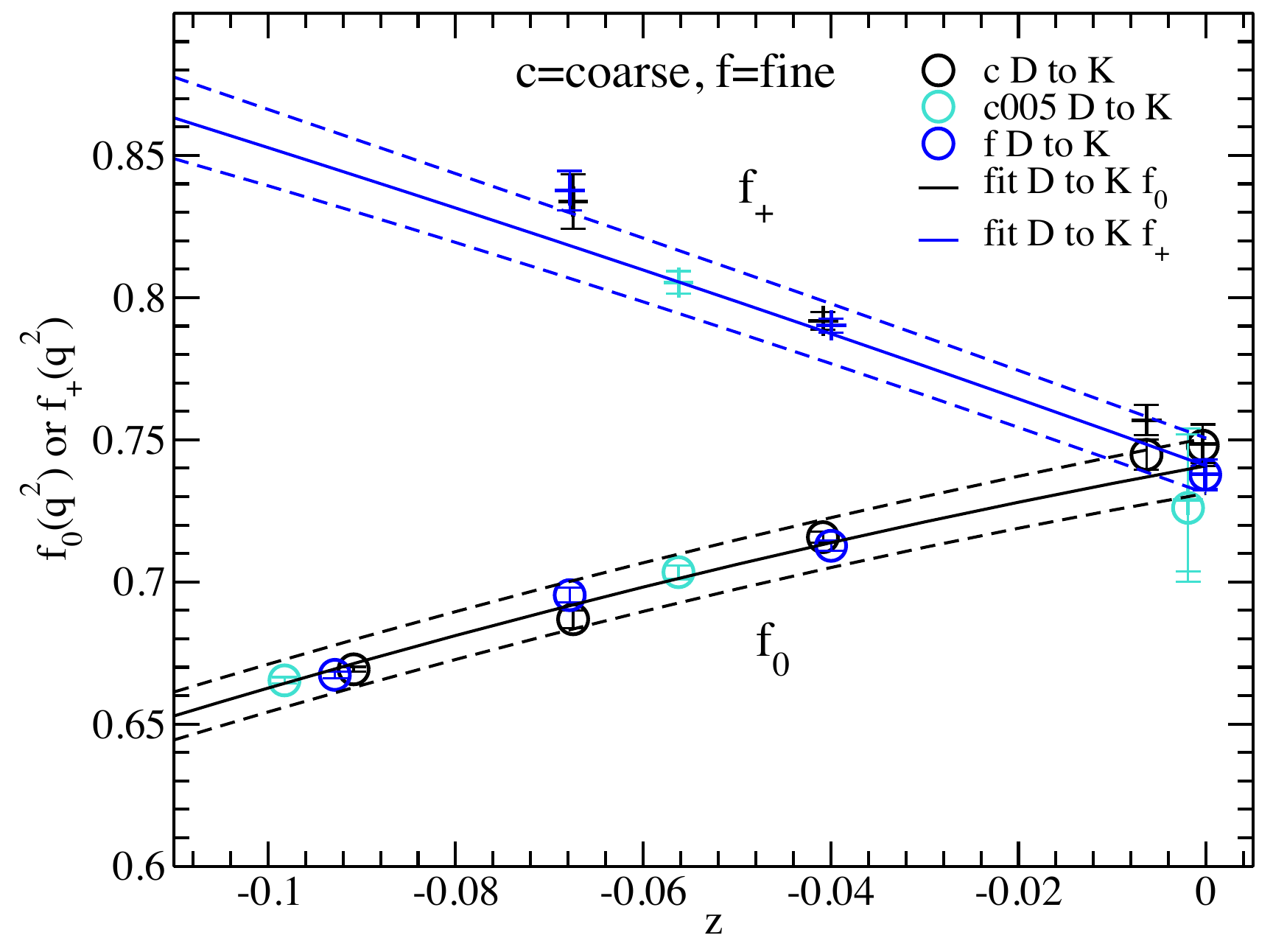}
\caption{$D\to K\ell\nu$ form factors and fit in $z$-space.}
\label{fig:zfit}
\end{figure}

\begin{figure}
\centering
\includegraphics[width=0.99\textwidth]{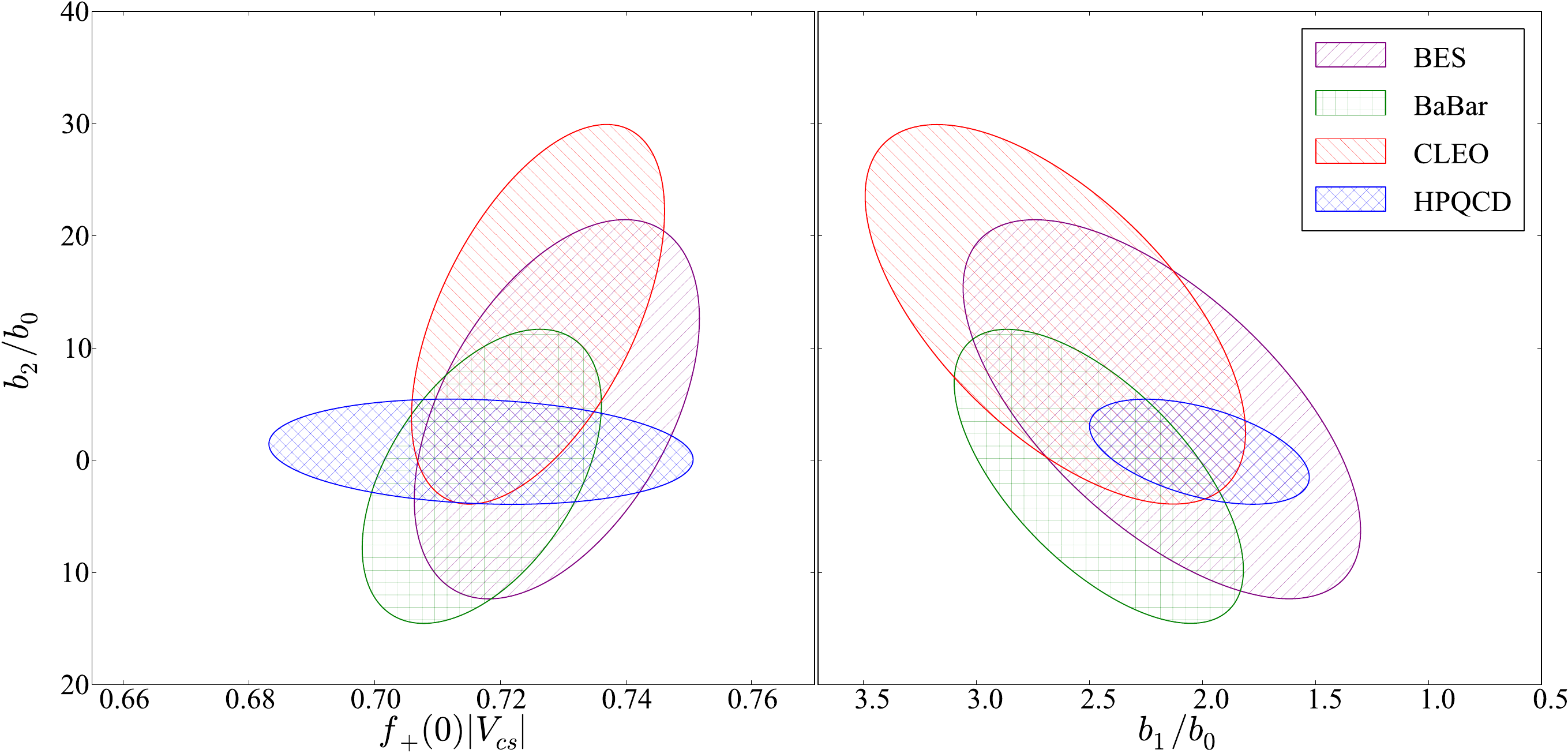}
\caption{To compare the shape of the $D \to K\ell\nu$ form factors calculated
in lattice QCD with experiment we use exactly the same $z$ expansion as the
experimental  groups~\cite{CLEO, BaBar, Belle, BES} (a more complicated
outer function than just a simple pole, and a specific choice of
$t_0$ in Equations~\ref{eq:zpole},~\ref{eq:q2toz}). 
Shown here are the $1\sigma$ ellipse contours of the
fit results from Eq.~\ref{eq:zexpansion} for $f_+(0)|V_{cs}|$ and $b_1/b_0$ against $b_2/b_0$. The
agreement is very good. $|V_{cs}|=0.963(5)_{\mathrm{expt}}(14)_{\mathrm{lattice}}$ was
used here for normalisation~\cite{HPQCD13}.
}
\label{fig:ellipse}
\end{figure}

It is beneficial to do the continuum and chiral extrapolation in
$z$-space instead of $q^2$-space. In $z$-space the semileptonic region 
is inside the unit circle, formed by the region with poles and cut -- 
see Fig.~\ref{fig:q2toz}. A simple conversion from $q^2$ to
$z$ is done as follows: First remove the poles
\begin{equation}
\tilde{f}_0^{D \to K}(q^2)=\Bigg(1-\frac{q^2}{M^2_{D^\ast_{s0}}}\Bigg)f_0^{D \to K}(q^2),\;\;
\tilde{f}_+^{D \to K}(q^2)=\Bigg(1-\frac{q^2}{M^2_{D^\ast_s}}\Bigg)f_+^{D \to K}(q^2),
\label{eq:zpole}
\end{equation}
then convert to z variable
\begin{equation}
z(q^2)=\frac{\sqrt{t_+-q^2}-\sqrt{t_+-t_0}}{\sqrt{t_+-q^2}+\sqrt{t_+-t_0}},\;
t_+=(M_D+M_K)^2,
\label{eq:q2toz}
\end{equation}
see e.g.~\cite{zconversion}. Remembering the constraint $f_+(0)=f_0(0)$ one can choose $t_0=0$. The
lattice results plotted in Fig.~\ref{fig:HPQCDFF} as a function of $q^2$ are
shown in Fig.~\ref{fig:zfit} as a function of $z$, which makes the
advantage of working in $z$-space very clear. The results from
different lattice ensembles are then fit as power series in z:
\begin{equation}
\tilde{f}_0^{D \to K}(z)=\sum_{n\geq 0}c_n(a)z^n,\; \tilde{f}_+^{D \to K}(z)=\sum_{n\geq 0}b_n(a)z^n,\;
c_0 = b_0.
\label{eq:zexpansion}
\end{equation}
Note that the fit parameters depend on lattice spacing and quark masses.
In the end one takes $a=0$ and $m_q=m_q^{\mathrm{phys}}$ to get the
result in the continuum and at physical light quark masses. Comparison
with experiment of the parameters from the $D \to K\ell\nu$ fit that 
determine the form factor shape is shown in  Fig.~\ref{fig:ellipse}.

\section{$|V_{cs}|$ and $|V_{cd}|$}
\label{CKMelements}

\begin{figure}
\centering
\includegraphics[width=0.9\textwidth]{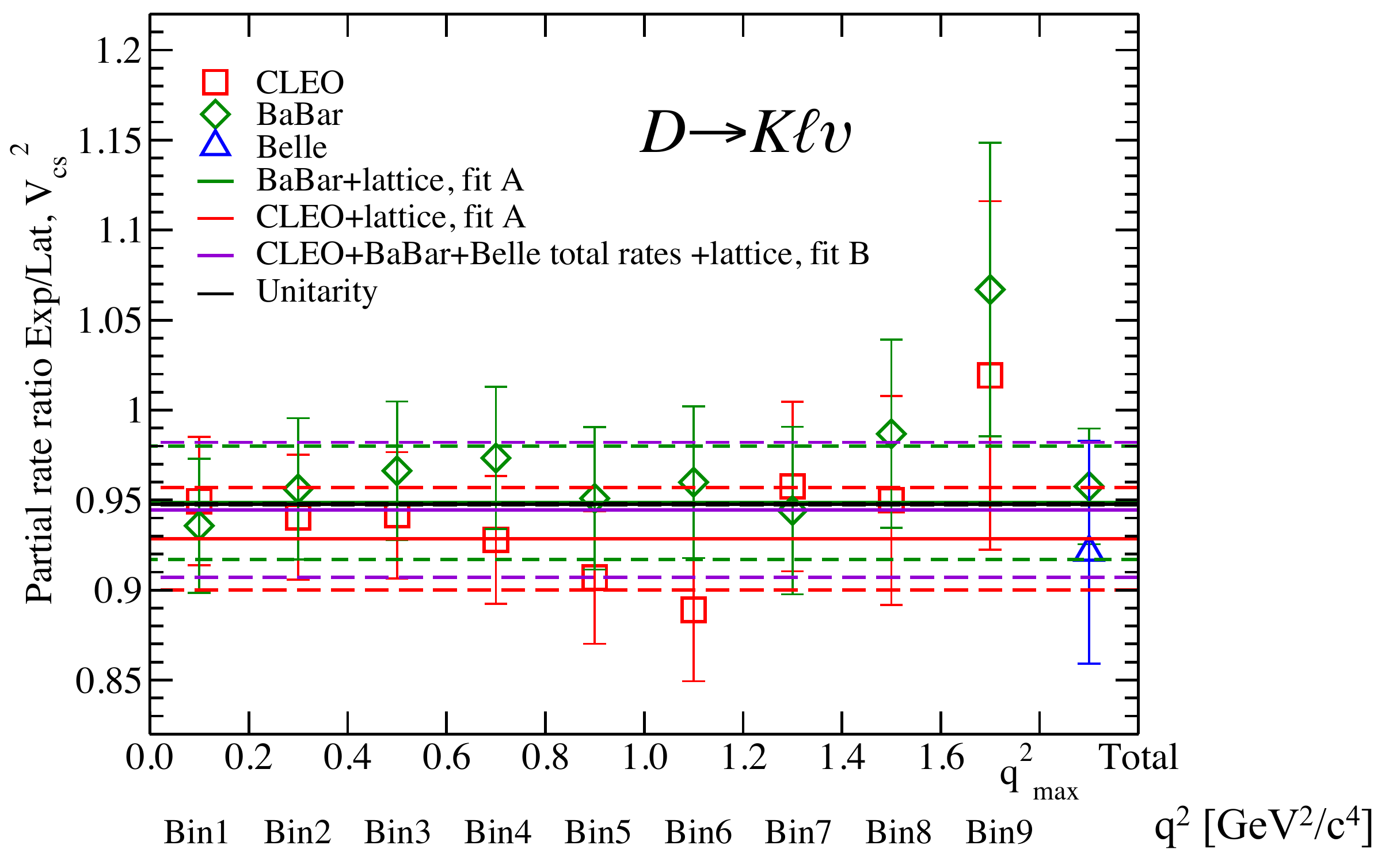}
\caption{$|V_{cs}|$ extracted from $D \to K\ell\nu$ decay using all
experimental $q^2$ bins.}
\label{fig:Vcsbin}
\end{figure}

\begin{figure}
\centering
\includegraphics[width=0.99\textwidth]{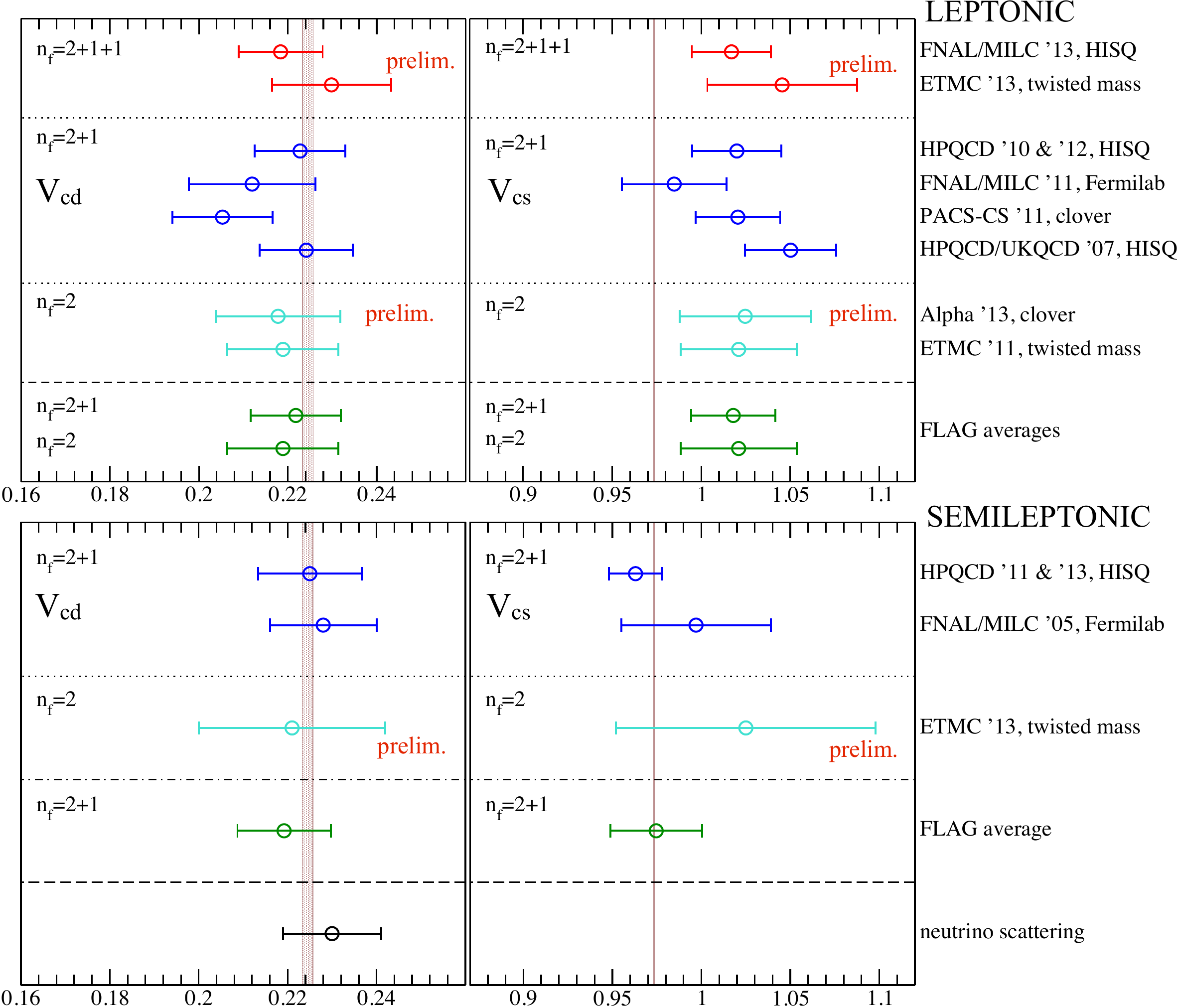}
\caption{Summary of CKM elements. Top row from left to right:
$|V_{cd}|$ and $|V_{cs}|$ from leptonic decays. Bottom row from left to right:
$|V_{cd}|$ and $|V_{cs}|$ from semileptonic decays. Vertical error
bands show the unitarity value for reference. The best values using
the latest lattice results (most modern lattice configurations with
$n_f=2+1+1$, and smallest errors) are: 
$V_{cd}~\textrm{(leptonic)} = 0.2184 \pm 0.009_{\mathrm{expt}} {}^{+0.0008}_{-0.0016} |_{\mathrm{lattice}}$
and $V_{cs}~\textrm{(leptonic)} = 1.017 \pm 0.02_{\mathrm{expt}}
{}^{+0.002}_{-0.007} |_{\mathrm{lattice}}$, taking decay donstants
from ~\cite{FNAL_MILC13} (FNAL/MILC '13 in Fig.~\ref{fig:fDfDs});
$V_{cd}~\textrm{(semileptonic)} = 0.225(6)_{\mathrm{expt}}(10)_{\mathrm{lattice}}$
from~\cite{HPQCD11} and
$V_{cs}~\textrm{(semileptonic)} = 0.963(5)_{\mathrm{expt}}(14)_{\mathrm{lattice}}$
from~\cite{HPQCD13}. Experimental averages used here are (taken from~\cite{FLAG}):
leptonic decays: $f_D|V_{cd}| = 46.40(1.98)$~MeV and $f_{D_s} |V_{cs}| = 253.1(5.3)$~MeV~\cite{RosnerStone};
semileptonic decays: $f^{D \to\pi}_{+}(0)|V_{cd}| = 0.146(3)$, $f^{D \to K}_{+}(0)|V_{cs}| = 0.728(5)$~\cite{HFAG}. 
The latest experimental results (2012 or after) are not included. \cite{HPQCD13} is the only calculation so far to
use all experimental $q^2$ bins to extract a CKM element from a
semileptonic decay.
}
\label{fig:VcdVcs}
\end{figure}

Now we have the needed input, decay constants and form factors, from lattice
QCD to determine CKM elements from leptonic and semileptonic decays.
In the case of  a semileptonic decay, we can integrate the form factor
calculated in lattice QCD over the experimental $q^2$ bins and determine
the CKM element from each bin: the experimental result divided by the lattice
result for a given bin is $V_{cs}^2$ (or $V_{cd}^2$). This is shown in
Fig.~\ref{fig:Vcsbin}. One can then do a weighted average fit to these values, 
including bin to bin correlations. 
This is more accurate compared to earlier calculations that
extracted CKM elements from experimental knowledge of $|f_+(0)|^2|V_{cs}|^2$ 
(or $|f_+(0)|^2|V_{cd}|^2$) and a lattice determination of the form factor at $q^2=0$,
since this uses more information.

The current status of $V_{cd}$ and $V_{cs}$ from leptonic and semileptonic decays
is shown in Fig.~\ref{fig:VcdVcs}. The tags are
the same as for the decay constants in Section~\ref{decayconst}: the
name of the lattice group, and the fermion discretization and number of
sea quark flavors that were used in the calculation. Note that the
experimental averages used here to calculate the CKM elements are from
2012~\cite{RosnerStone,HFAG,PDG} -- more recent experimental results
have not been included. The vertical lines in the plots show the
unitarity value. Leptonic decays tend to give a higher value for
$V_{cs}$ than the unitarity value, but note that all data points in
the plot would shift to left or right, if the experimental average
changed. All lattice results agree with each other very well, and the
semileptonic determination of $V_{cs}$ and both leptonic and
semileptonic determinations of $V_{cd}$ agree with the assumption of
CKM matrix unitarity. For comparison, averages from Flavor
Lattice Averaging Group (FLAG)~\cite{FLAG} are also shown in the plots,
as well as the result for $V_{cd}$ from neutrino scattering experiments~\cite{PDG}.
The lattice results are from
\cite{ETMC11,FNAL_MILC05,HPQCD_UKQCD07,HPQCD11,PACS-CS11,FNAL_MILC11,HPQCD12,HPQCD13,ETMC13,Alpha13,FNAL_MILC13}.

\section{Summary}

Precision tests of Standard Model and searches for new physics can 
be done by extracting CKM elements from $D$ and $D_s$ meson leptonic 
and semileptonic decays. In addition to precise experimental results 
input from theory is also needed: decay constants $f_D$, $f_{D_s}$, 
and form factors for $D \to K\ell\nu$, $D \to \pi\ell\nu$, $D_s \to K\ell\nu$. 
These can be calculated in lattice QCD, and the current best results for
the decay constants are listed in Fig.~\ref{fig:fDfDs} along with the 
corresponding CKM elements in Fig.~\ref{fig:VcdVcs}. 
Results from independent lattice calculations show good agreement, and 
the extracted $|V_{cd}|$ and $|V_{cs}|$ are in agreement with CKM matrix 
unitarity.  We also compare the shape of the form factors from lattice 
QCD with experimental results and find good agreement.

\Acknowledgements

I am grateful to Christine Davies, Gordon Donald and Andreas Kronfeld
for useful discussions. Thank you to the organisers for an excellent meeting!

\end{document}